\documentclass[letterpaper,pre,showpacs,12pt, superscriptaddress]{revtex4-1}
\usepackage{color}
\usepackage{graphicx}
\usepackage{dcolumn}

\newcommand{\ie} {{i.e., }}

\newcommand{\pc} {p_{\rm c}}
\newcommand{\pd} {\partial}

\newcommand{\xhat} {\hat{\bf x}}

\newcommand{\tw}[1]{{\color{blue}[$[_{\bf TW}$ #1]]}}
\newcommand{\no}[1]{{\color{magenta}[$[_{\bf NO}$ #1]] }}

\newcommand{\omitt}[1]{} 
\newcommand{\lessthanorabout}
           {\mathrel{\raise.3ex\hbox{$<$\kern-.75em\lower1ex\hbox{$\sim$}}}}
\newcommand{\greaterthanorabout}
           {\mathrel{\raise.3ex\hbox{$>$\kern-.75em\lower1ex\hbox{$\sim$}}}}

\begin{document}

\title{Anomalously fast kinetics of lipid monolayer buckling}

\author{Naomi Oppenheimer}
\affiliation{James Franck Institute,
University of Chicago, Chicago, Illinois 60637}
\email{naomiop@uchicago.edu}

\author{Haim Diamant}
\affiliation{Raymond \& Beverky Sackler School of Chemistry, Tel Aviv University, Tel Aviv 69978, Israel}

\author{Thomas A. Witten}
\affiliation{James Franck Institute,
University of Chicago, Chicago, Illinois 60637}

\date{\today}

\begin{abstract}
We re-examine previous observations of folding kinetics of compressed lipid monolayers in light of the accepted mechanical buckling mechanism recently proposed [L.\ Pocivavsek {\it et al.},
{\it Soft Matter}, 2008, {\bf 4}, 2019].  Using simple models, we set conservative limits on a) the energy released in the mechanical buckling process and b) the kinetic energy entailed by the observed folding motion.  These limits imply a kinetic energy at least thirty times greater than the energy supplied by the buckling instability. We discuss possible extensions of the accepted picture that might resolve this discrepancy.
\end{abstract}


\maketitle

\section{Introduction}
\label{sec_intro}
\begin{figure}[tbh]
\vspace{0.7cm}
\centerline{\resizebox{0.45\textwidth}{!}{\includegraphics{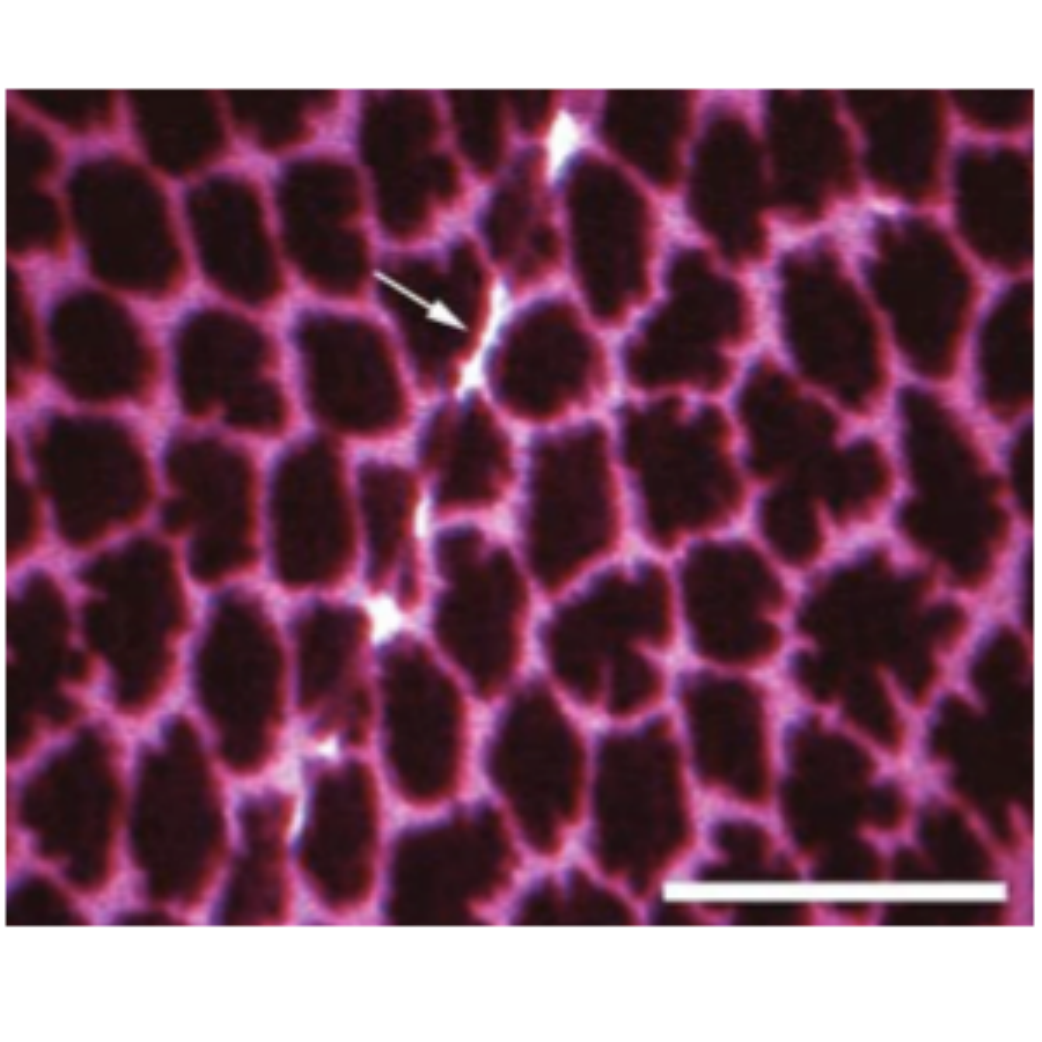}}}
\caption{The bright line is a fold created in a compressed  mixture of DPPG:POPG lipids. Dark domains are the condensed phase surrounded by a liquid phase in purple. Bar indicates 50 $\mu$m. 
Figure taken from Ref.~\cite{jpc06}.}
\label{fig_buckle}
\end{figure}

When a flat thin sheet
of material is subjected to increasing pressure, it eventually
buckles, crumples, or cracks.  By buckling, the sheet releases stress
over wavelengths much larger than its thickness.  During the past
years there has been increasing interest in phenomena, such as
crumpling, where the deformation goes from an initial uniform state to
a localized region occupying an arbitrarily small fraction of the
sample \cite{focusing}. 
Monolayers of surface-active molecules (surfactants), adsorbed on a
liquid, are found in many systems containing water--air or water--oil
interfaces \cite{Birdi}. Such monolayers exhibit a rich variety of collapsed structures
under lateral compression, most of which occur on an intermediate
scale between the macroscopic one and the molecular thickness of the
layer.  Fluid monolayers collapse into disks, tubes, or pearls-on-string structures, depending on spontaneous curvature and charge of the lipid monolayer \cite{KYLeeReview}.
Like elastic sheets, many solid-like monolayers crack
\cite{break1,break2} or buckle \cite{buckle} under pressure, yet other
solid-like monolayers fail by abrupt buckling into straight,
micron-wide folds
\cite{Lee98,Gopal01,Knobler02,Bruinsma,Fischer,Longo,jpc06,LukaCerda, Kwan} (See Fig.~\ref{fig_buckle} for an example).  In
addition, liquid-like monolayers may form micron-scale vesicular
objects of various shapes \cite{Gopal01,Nguyen,Hatta07} or giant
convoluted folds \cite{Knobler02,Bruinsma}. This type of folding is believed to be driven by the interfacial energy gained from the contact across the two sides of the fold \cite{Bruinsma}. Thus, the failure of
surfactant monolayers under lateral pressure displays distinctive
mechanical behaviors, which crucially depend on the in-plane rigidity
\cite{Gopal01,Gopal06,sm} and viscoelasticity \cite{Bruinsma}. 
Lipid monolayers and bilayers are the material of choice for spatial
partitioning in living matter, such as cells and compartments within
them.  These partitions are often observed to fold and wrinkle under
stress.  In particular, lipid monolayers that model the expanding and
contracting sacs in an animal's lung exhibit the abrupt folding
signature described in Refs. \cite{lung1,lung2,lung3}.

This abrupt buckling is particularly puzzling.  The abrupt relaxation
motions or ``jerks" have been extensively characterized for a particular system \cite{jpc06}, but their distinctive length and time scales have
not been explained.  Why do the jerks occur over a time scale of about
0.1 seconds---far from any molecular or apparatus time-scales? Why
does the motion stop suddenly, before the driving stress is relaxed?
Why do the jerks show a robust characteristic displacement of a few
microns?

An intriguing hypothesis to explain this micron length scale was
recently proposed by Pocivavsek et al \cite{LukaCerda, KYLeeReview}.  These authors
recalled that any thin sheet under compression on a liquid substrate
buckles at a well-defined wavelength $\lambda$.  For lipid monolayers
the expected value of $\lambda$ lies in the micron range.  Thus it is
of the same order as the characteristic displacement of the jerk
relaxations.  These authors also noted that the incipient wrinkles at
wavelength $\lambda$ are unstable against folding, in which the excess
wrinkled material from throughout the sample is concentrated into a
single loop or fold.

This paper aims to account for the dynamical features of the jerk motion.  In Section \ref{sec_system} we argue that the monolayer may be viewed as a broad, thin slab that translates almost rigidly during the jerk motion.  Then in Section \ref{sec_dynamics} we describe how such a slab should respond to horizontal forcing, accounting for progressive viscous entrainment of the fluid subphase.  Using this result, combined with experimental observations, we set a lower bound on the kinetic energy of the jerks.  We then survey the possible forces that might give rise to this kinetic energy, notably the energy released by folding in the mechanism of Ref. \cite{LukaCerda}.  Even upper limits of this folding energy are far less than this observed kinetic energy. In Section \ref{sec_discuss} we discuss effects that might account for this kinetic energy.

\section{system}
\label{sec_system}

For definiteness we focus our study on the jerks analyzed in Ref. \cite{jpc06} and re-examined in Ref. \cite{LukaCerda}.  We begin by reviewing the parameters of this system. The monolayer was a 7:3 mixture of dipalmitoylphosphatidylcholine (DPPC) and dioleoylphosphatidylglycerol (POPG) spread on pure water at room temperature in a $15$-cm-long Langmuir trough and viewed in a microscope in a 100-micron-wide field of view at conventional video frame rates. It is compressed at 0.1 mm/sec to a nominal pressure of about 70 mN/meter before viewing.  The measured pressures are consistent with the slight overpressure needed for folding \cite{LukaCerda,TWWilhelmy}.  In these conditions this material microphase separates into a biphasic foam-like pattern of compact patches separated by narrow strips of different composition, visualized by a dilute fluorescent additive. The patches are 15-25 microns in size.  The monolayer behaves mechanically as a solid, not a fluid; the jerks move in the direction of the Langmuir barrier motion.  Rheological measurements in similar systems showed stress relaxation times of order $10$--$10^2$ sec.\cite{Mohwald,Bruinsma}. The jerks vary statistically in their net displacement $\Delta$ and their duration $t$. No statistical correlations between jerks were observed. The displacements $\Delta$ vary from a minimum of about 1.2 microns to several times larger, with an average of 2 microns \cite{jpc06}. The durations $t$ vary from a minimum of about $0.09$ sec to a few times longer, where the average is 0.12 sec.  It has been suggested \cite{jpc06} that the larger jerks are cascades of elementary jerks, with this average $\Delta$ and $t$.  In the estimates below, we use these average $\Delta$ and $t$ values.

We may simplify our description of the monolayer using three further features. 
First, the monolayer may be assumed to translate {\it rigidly} outside the folding region over distances of several cm.  It thus entrains substrate fluid over these distances.  In principle compressibility could invalidate this assumption. Compressibility implies that an initial imbalance of membrane stress produces a compressional wave whose speed is given by, $c=\sqrt{Y/\rho_s}$, where $Y$ is the two-dimensional uniaxial compression modulus and $\rho_s$ is the mass density per unit area. If forces are applied on a timescale $t$, the resulting compression or expansion is confined to distances $L \lessthanorabout c t$.  Conversely, the compression is negligible and the body moves rigidly if its size $L$ is smaller than about $c t$. Thus to show that our monolayer translates rigidly, we must establish that $c$ is sufficiently large.

For our case the density $\rho_s$ is the density of the material to be accelerated in propagating the wave. Since the monolayer entrains substrate fluid as it moves, the density must take account of this subphase.  We may find a lower limit on this $c$ by using a lower limit to the modulus $Y$ and an upper limit for the surface density $\rho_s$.  Accordingly we estimate $Y$ by neglecting any compressional effects of the subphase.  Measurements of the monolayer compression modulus range from 0.1 N/m to several N/m \cite{Gopal06, compressibility1, compressibility2}.  
\omitt{For the system in question, the derivative of the surface pressure gives a modulus of several times the osmotic pressure. } We use the conservative estimate of $Y > 0.1$ N/m below.  As for the surface density $\rho_s$ we find an upper limit by including all the water that might be entrained. As discussed in Section \ref{sec_dynamics}, an upper limit on the entrained density $\rho_s$ is given by $\rho_s < 2\rho \sqrt{\nu t}$, where $\rho$ ($10^3$ kg/m$^3$) is the density of the water subphase, and $\nu$ ($10^{-6}$ m$^2$/sec) is its kinematic viscosity.  Thus in the jerk time of 0.12 sec, a local compression can propagate a distance  $c~ t > 0.05$ m, i.e. even by our minimal assumption, information from the folding region has reached about half the sample (5 cm) within the jerk time. Thus we expect only minor effects from compressibility of the sheet. In the calculations to follow we will assume $L= 5$ cm is the size of the moving sheet.

A second feature of the monolayer is that its compressibility has negligible effect on the energetics of folding.  As explained below, the over-pressure $p$ released in the wrinkle-to-fold transition \cite{LukaCerda} in this system must be smaller than $6 \times 10^{-5}$ mN/m.
Given the large lower-bound modulus $Y$ above, the compressive displacement $\Delta_{\rm com}$ in a system of length $L$ is smaller than $L p/ Y < 0.03$ micron. \omitt{from above p= 6E-8, Y= .1, so strain p/Y = 6E-7=.6E-6. displacement \Delta_{
\rm com} = L p/Y = .05*.6 E-6 Meters = .03 micron. NO:  I just compared $E \sim Y_{3D} A/L \Delta_{\rm com}^2= Y_{3D} d \omega /L \Delta_{\rm com}^2 = Y_{2D} \omega/L \Delta_{\rm com}^2$, and $E\sim p\Delta_c \omega$, to obtain $\Delta_{\rm com}$. } It is thus much smaller than the observed displacements $\Delta$.  Likewise, the compressive energy released is a small fraction ($p \Delta_{\rm com} ~/~ p \Delta$) of the folding energy.

\begin{figure}[tbh]
\vspace{0.7cm}
\centerline{\resizebox{0.48\textwidth}{!}{\includegraphics{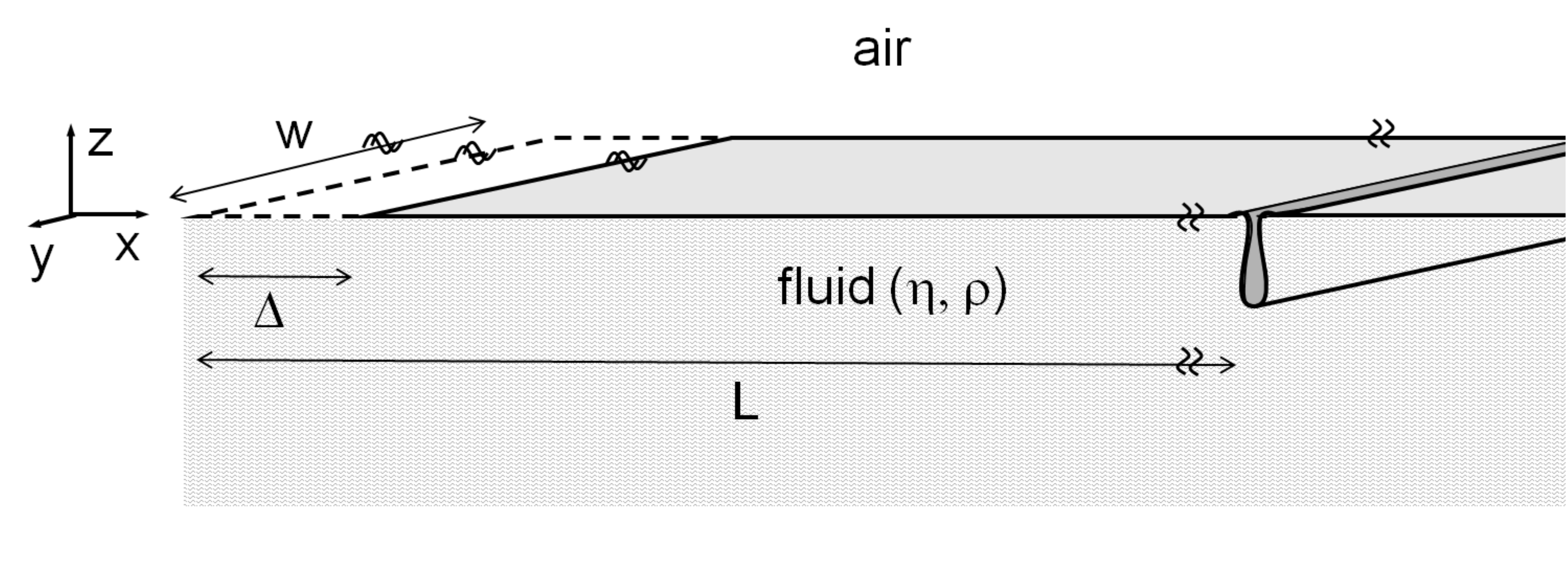}}}
\caption{Schematic view of the system and its parameterization.}
\label{fig_scheme}
\end{figure}

A third feature of these monolayers is that the
longitudinal propagation of the fold tips is much faster than the
transverse folding and its accompanying jerk. Within the temporal
resolution of Ref.\ \cite{jpc06} ($\sim 0.03$ s) the fold traverses
the field of view ($\sim 150$ $\mu$m) instantaneously; this sets a
lower bound of $\sim 5$ mm/s for its speed. In the system of Ref.\
\cite{Fischer} two types of longitudinal folds were observed, the
slower of which propagated at $\sim 10$ mm/s. Comparing these values
with the characteristic translation velocity, $10$ $\mu$m/s \cite{jpc06}, we see
that the requirement is safely fulfilled.  These features
allow us to consider a simplified two-dimensional problem of a thin
elastic sheet moving over a semi-infinite viscous liquid (Fig.\ 
\ref{fig_scheme}).

Current understanding \cite{LukaCerda, KYLeeReview} attributes the jerks to the mechanical buckling or wrinkling instability of any elastic sheet that is floating on a liquid and is under compression \cite{Milner}. Above a threshold pressure $p_c$, the sheet distorts, at a cost of bending energy.  The wrinkles also produce a net upward and downward displacement of the liquid, thus increasing its gravitational energy. Taking account of these costs, one finds a threshold pressure given by $p_c = 2 \sqrt{B \rho g}$, where $B$ is the bending stiffness and $g$ is the acceleration of gravity.  The predicted buckling wavelength $\lambda$ is given by  $\lambda = 2\pi [B/(\rho g)]^{1/4}$.  This buckling is unstable, leading to a release of the overpressure $p_c$ over a displacement $\Delta_f$ comparable to $\lambda$.  Though $p_c$ is too small to measure directly in our system,  it can be estimated either using typical values of $B$ or by inferring $B$ from the observed jerk displacement.  Bending stiffness for lipid monolayers like our system lie in the range $B \lessthanorabout 10^{-19}$ J. \cite{bendinmoduli}. This $B$ value implies an upper bound for $p_c$: $p_c < 6 \times 10^{-5}$ mN/m.  If instead we infer $p_c$ from the observed jerk displacements via  $p_c = \rho g \lambda^2/(2\pi^2) = \rho g \Delta_f{}^2/(2\pi^2) $, we obtain values of $2 \times 10^{-6}$ mN/m, i.e. 30 times smaller than our upper bound \cite{bending_factor}. In what follows we will use the more conservative ``upper-bound" value of $p_c$, namely $6\times 10^{-5}$ mN/meter.
There are many forces in the system that exceed this value - the total pressure from the trough, drift flows, flows from faraway jerks, or residual flows from previous jerks. 
However, these larger pressures lack the central feature needed to explain jerks. The jerks clearly result from an instability, in which a small displacement $\Delta$ leads to an increasingly {\it unbalanced} pressure. 

To summarize the above discussion, the sheet is assumed to deform in the $x$--$z$ plane while remaining
uniform along the $y$ axis. It is laterally
compressed by a two-dimensional, uniform pressure $p$.  (In the
experimental system $p$ is given by the actual pressure exerted at the
boundaries minus the surface tension of the liquid.)  Prior to
instability the monolayer remains flat ($h=0$) and responds to compression by slightly
decreasing its actual length. When the pressure exceeds $p_c$, the monolayer wrinkles or folds out of the $x$--$y$ plane ($h \neq 0$). Since the pressure $p_c$ produces negligible elastic compression, the total length $L$ is fixed, and the displacement $\Delta$ along the $x$ axis is fully accounted for by the
wrinkles or folds.  
The underlying liquid, having viscosity $\eta$ and mass density
$\rho$, occupies the region $z<h$.

%

\section{Dynamics}
\label{sec_dynamics}

\subsection{Constant Pressure}
\label{sec_constant}

As noted above, the motion of the sheet depends strongly on the viscous entrainment of the fluid beneath it.  In this section we determine the motion taking account of this entrainment.  We begin with a simplified situation.
Suppose we control the surface pressure and increase it above the
instability pressure, $\pc$. We further suppose that as the sheet
begins to fold, the buckled region is decreasing the pressure by a constant amount $p$. At this moment the forces on the sheet are unbalanced and it begins to accelerate. In all but the small folding region the sheet is horizontal and its motion is a pure horizontal translation. The flat sheet adjacent to the fold will
translate laterally, thus creating a velocity profile $v(z)\xhat$ in
the underlying liquid. On larger scales this flow produces circulation of fluid in the sample. We ignore such large scale gestures, since these only increase the kinetic energy and the discrepancy we are discussing. The resulting viscous drag decreases with time,
as the velocity profile penetrates deeper into the liquid and its
gradient becomes less sharp. We shall assume that the sheet's velocity is always equal to that of the water immediately beneath it.  That is, the fluid obeys a no-slip boundary condition. We examine such assumptions in Section \ref{sec_discuss} below. We may neglect the mass of the sheet relative to the much greater mass of the entrained fluid. Thus the problem is simplified to that of finding the velocity of a
half infinite fluid due to a rigid surface moving with a constant lateral
pressure.

The equation of motion for the fluid is:
\begin{equation}
  \dot{v} = \nu\pd_{zz} v,
\label{fluid_motion}
\end{equation}
where $\nu$ is the kinematic viscosity given by the ratio of viscosity and density, 
$\nu=\eta/\rho$, and a dot denotes derivative with respect to time.
The constant stress at $z=0$ implies a boundary condition of the form
\begin{equation}
	\partial_z v|_{z=0}=-\frac{p}{L\eta}.
	\label{pressure}
\end{equation}
Accompanying it are the condition of vanishing flow far away from the interface,
\begin{equation}
  v(z\rightarrow-\infty,t) = 0,
\label{bc2}
\end{equation}
and initial conditions of stationarity,
\begin{equation}
  \Delta(t=0) = \dot{\Delta}(t=0) = v(z,t=0) = 0.
\label{bc3}
\end{equation}
Eqs (\ref{fluid_motion}) and (\ref{pressure}) may be simplified by defining the Laplace transform $\tilde v(\omega)$ by $\tilde v(\omega) \equiv \int_0^\infty v(t) e^{-\omega t} dt$:
\begin{equation}
\omega \tilde{v}(z,\omega)=\nu \partial_{zz}\tilde{v}(z,\omega), \ \ \   
\partial_z\tilde{v}|_{z=0}=-\frac{p}{\eta L \omega}. 
\label{laplace_motion}
\end{equation}
The solution in Laplace space is
\begin{equation}
\tilde{v}(z,\omega)=\frac{p\sqrt{\nu}}{\eta L \omega^{3/2}} ~\exp(\sqrt{\omega/\nu}~ z),
\label{solution_laplace}
\end{equation}
and in the time domain,
\begin{equation}
v(z,t)=2 p(\nu t/(\pi L^2\eta^2))^{1/2}[e^{-\alpha^2}+\alpha({\rm Erf}(\alpha)+1)], 
\  \ \alpha\equiv z/(2\sqrt{\nu t})
\label{solution_real}
\end{equation}
%
The displacement of the sheet is given by integrating the velocity on 
the surface, $\Delta=\int_{0}^t v(z=0,t')dt'$,
\begin{equation}
  \Delta = \frac{4}{3\sqrt{\pi}}\frac{p\sqrt{\nu}}{L\eta} ~t^{3/2} 
\label{power32}
\end{equation}
Since the accelerated mass is constantly increasing in time, the acceleration $\dot v$ decreases with time as $t^{-1/2}$.  This leads to the unusual "jerky" increase of the velocity $v(t)$.  We note that this predicted motion arises from basic hydrodynamics; it does not depend on the source of the pressure $p$.  We note also that a fixed fraction $\simeq 0.64$ of the input power $p \dot \Delta$ goes into kinetic energy; the remainder goes into viscous dissipation.
\omitt{ {\tt numbers to be checked; perhaps exact expression can be quoted as a footnote.} \no{\tt Agree, don't think we need to add exact expression} 
$$
1 - \frac{8-5 \sqrt{2}+6 \left(\sqrt{2}-2\right) \sqrt{\pi }+3 \sqrt{2} \pi }{4 \pi }
$$}
\subsection{General Pressure}

Buckling relaxes stress at the tip of the fold; we would therefore expect 
the driving pressure to depend on the displacement. The equation for $\Delta(t)$ can readily be generalized to include such a $p(\Delta)$.  We discuss it here for completeness.
\begin{equation}
  L \sigma \ddot{\Delta} = p(\Delta) - L\eta\pd_z \left . v\right |_{z=0},
\label{motion1}
\end{equation}
where the first term on the right is the driving stress, the last term is the stress from the fluid, 
and $\sigma$ is the two-dimensional mass density of the sheet. 
For the fluid, Eq. (\ref{fluid_motion}) still applies, and the two 
are supplemented by a no-slip boundary condition at the interface, 
\begin{equation}
  v(z=0,t) = \dot{\Delta},
\label{bc1}
\end{equation}
a condition of vanishing flow far from the surface (Eq. \ref{bc2}), 
and initial conditions of stationarity (Eq. \ref{bc3}). 

We proceed by eliminating $v$ to obtain an equation for the sheet alone. 
The velocity Green function of the liquid is given by the boundary 
condition $v(z=0,t)=\delta(t)$ and Eqs. (\ref{bc2},\ref{bc3}).
%
%
%
Explicitly,
\begin{equation}
	G(z,t,t')=\frac{1}{2\sqrt{\pi \nu}}\frac{z}{(t-t')^{3/2}}\exp\left(-\frac{z^2}{4\nu (t-t')}\right).
\label{G}
\end{equation}

The general solution with the no-slip boundary condition of Eq.
(\ref{bc1}) is obtained simply by integration, $v(z,t)=\int_0^t
G(t-t')\dot{\Delta}(t')dt'=\int_0^t G(t')\dot{\Delta}(t-t')dt'$.
Inserting this expression in Eq.\ (\ref{motion1}) and integrating by parts, produces an equation for the sheet alone:
\begin{equation}
  L \sigma \ddot{\Delta} = p(\Delta) -
  \frac{L\eta}{\sqrt{\pi\nu}} \int_0^t dt' \frac{\ddot{\Delta}(t-t')}
  {\sqrt{t'}}.
\label{power32}
\end{equation}
As before, we may neglect the inertial term on the left hand side. For a constant pressure, Eq. (\ref{power32}) 
gives the same displacement as Eq.(\ref{solution_real}).
\omitt{\tw{\tt Oddly the current value of p seems to be determined by the past values of $\ddot \Delta$.  Since we think of the force as causing the motion, this seems backwards.  }. \no{\tt Is it clearer if we write the equation in Laplace space? $\omega^2 L \sigma \tilde{\Delta} = \tilde{p}(\Delta)-L \eta \omega^{3/2}\tilde{\Delta}/\sqrt{\nu}$.} }

\subsection{Application to observed jerks}
\label{sec_application}

We may now compare the observed jerk motion with the motion expected from the folding forces. It is straightforward to solve Eq. (\ref{power32}) using the predicted (quadratic) $p(\Delta)$ \cite{Diamant2011}.  The resulting motion proves to be much slower than the observed jerks. We may quantify this discrepancy in two ways.  First we ask what constant pressure $p$ would be required to give the observed displacements $\Delta$ in the observed time $t$.  Then we determine the observed kinetic energy and compare it to the energy available from folding.

To find the pressure required to produce the observed jerks, we substitute $\Delta = 2$ microns and $t = 0.12$ sec into Eq. (\ref{power32}), to obtain $p =3.2 \times 10^{-3} $ mN/m. This is some 50 times larger than the upper-bound folding pressure $p_c = 6 \times 10^{-5}$ mN/m obtained above. 
\omitt{\tw{\tt I think some numbers have to change in this section in light of some factors of 2 etc that you found.} \no{\tt I think these are fine, the problem and the factor 2 were in Eq. 13 which is yet to come, I changed some values there}. }

The same discrepancy emerges if we compare the observed kinetic energy $E_k$ with the work $W_p$ done by the pressure difference $p(\Delta)$. Both are proportional to the width $w$ of the jerking region. The pressure $p(\Delta)$ is always smaller than $p_c$ throughout the folding.  Thus $W_p/w < p_c~\Delta < 1.2 \times 10^{-13}$ N . Other things being equal, the kinetic energy $E_k$ for a given average velocity $\Delta/t$ is larger if $t$ is larger (since larger $t$ implies a greater entrained mass).\omitt{\tw{\tt the statement is only supposed to compare a given $\Delta(t)$ with a slowed down version $\Delta(a~ t)$, Maybe this is clearer with my addition.}}  Thus we may obtain a lower bound on the kinetic energy by limiting $t$ to the observed duration of $0.12$ sec. For the moment we simply use the constant-pressure solution of Eq.~(\ref{solution_real}) and use 
\begin{equation}
E_k/w = L~\frac 1 2 \rho \int dz~ v(z)^2 = 1.7 ~L \sqrt{\nu t}~ \left [\frac 1 2 ~\rho~ \left (\frac \Delta t \right )^2 \right]
\label{max_kinetic}
\end{equation}
We examine this estimate in Sec. \ref{sec_discuss}. For $L = 0.05 $m, $\Delta = 2$ microns and $t = 0.12$ sec, this gives $E_k/w = 4 \times 10^{-12}$ N---some 32 times our upper bound of the work $W_p$ supplied by folding.  
Thus even when one ignores the work that must go into viscous dissipation, the observed energy is far larger than what the folding energy ,$W_p$, can supply.  (Taking account of the dissipated energy, one recovers the factor-50 discrepancy quoted above.)

\section{Discussion}
\label{sec_discuss}

The arguments above indicate a worrisome discrepancy between the observed jerking motion and the mechanics of folding presumed to account for this motion.  The discrepancy is serious; it survives even when we use conservative bounds in our estimates.  In this section we survey possible ways to account for the discrepancy.  First we review possible flaws in our description of the kinetic energy and the folding forces.  Then we consider other forces that might account for the jerking motions.

Our estimate of the kinetic energy was a simplified one, but it gives a proper lower bound for a given displacement $\Delta$ and time $t$. We will proceed by reinforcing several of the assumptions.  
Firstly, we assumed a constant pressure, though the actual unbalanced pressure increases with time. However other choices would have led to a higher kinetic energy. 
We consider the effect of replacing our constant-pressure estimate by allowing the pressure to increase with time. In order to achieve the required $\Delta (=\int dt~ \dot \Delta)$ in the given time $t$ with a time-increasing pressure, we will necessarily reduce $\dot \Delta$ at early times and increase it at late times.  However, any shift of $\dot \Delta$ from earlier to later times has the effect of increasing the kinetic energy. To see this, we consider a small decrease of $\dot \Delta$ at time $t_<$ over a brief interval $\Delta t$. To maintain a fixed total displacement $\Delta$, we make an equal addition to $\dot \Delta$ at a later time $t_>$.  Any shift of $\dot \Delta$ from earlier to later times can be accomplished by repeating this process.  This perturbation of $\dot \Delta$ creates a corresponding perturbation of the fluid velocity at the final time $t$: we denote it by $\delta v(z, t)$.  We may then express the final kinetic energy  $E_k(t)$ in terms of this $\delta v$ and the initial profile $v_0(z, t)$ using the integral of Eq.~\ref{solution_real}.
\begin{eqnarray}
E_k(t)/w~  &&= L~\frac 1 2 \rho \int dz~ [v_0(z, t) + \delta v(z, t)]^2 \nonumber\\
&&= E_{0k}(t)/w + L \rho \int dz~ v_0(z, t)~ \delta v(z, t) ~~+ {\cal O} (\delta v^2)
\label{kineticdeltav}
\end{eqnarray}
The second perturbing term is necessarily positive provided the (positive) $v_0(z)$ profile is monotonic.  To see this we express $\delta v$ in terms of the Green function $G$ of Eq.~(\ref{G}): 
\begin{equation}
\delta v(z, t) = \delta~ \Delta t ~[-G(z, t-t_<) + G(z, t - t_>)].
\end{equation}
We may express any monotonic $v_0$ as a sum of positive step functions extending from 0 to some $Z$.  For a given step function, the contribution to $E_k$ is given by $\int_{-Z}^0 \delta v$.  Thus it suffices to show that this integral is positive.  In terms of the $G$ functions, this means $
\int_{-Z}^0 G(z, t-u)$ is an increasing function of $u$.  This may be verified explicitly using Eq.~(\ref{G}). Thus a small shift in the pressure profile from a constant one to an increasing one with the same  $\Delta$ and $t$ only increases the energy $E_k$. If further small shifts are added, the same reasoning implies that the $E_k$ again increases, provided the starting $v_0(z)$ remains monotonic in $z$.  We conclude that the constant-pressure $E_k$ of Eq.~(\ref{max_kinetic}) under-estimates $E_k$.

A second assumption that affects our estimate of the kinetic energy is the no-slip boundary condition between the sliding monolayer and the fluid beneath. Having a slip would cause less drag of the fluid and thus less kinetic energy. But in our case of hydrophilic
heads facing the water, there is no justification for a significant slip. 

A further effect that can potentially reduce the kinetic energy is the possibility that the motion is restricted in area, so that less fluid is entrained.  As noted above, this restriction can occur if the surface layer is compressible.  The expected compressibility is such that there could be a noticeable effect on the scale of the entire $15$ cm sample.  To account for this, we assumed that the motion is restricted to a range of only 5 cm.  On this scale we argued that any departure from rigid sliding of the sheet would be negligible. One possibility we did not consider is that the modulus of the sheet, $Y$, is smaller than observed in experiments (that dealt with sheets without any folds). The existence of folds would make the material weaker, softer. If that is indeed the case, the part of the sheet that moves rigidly due to jerking could be smaller. However, in order to resolve the discrepancy, the rigid moving part must be a few millimeters at most, which is not consistent with the statistics of observed jerks \cite{jpc06}. 

Another potential way to resolve this discrepancy is that we have {\it under}estimated the folding pressure $p$. As was noted in section \ref{sec_dynamics}, the creation of a fold reduces the pressure. Since folding must occur when the pressure exceeds $p_c=2\sqrt{B\rho g}$ no greater excess pressure can be sustained. This is true even if additional folds are present. Could it be that $p_c > 6 \times 10^{-5}$ mN/m? This would imply that we have greatly underestimated the bending rigidity $B$. The local structure of the monolayer might imply a bending rigidity greater than is usually observed due to texture in the sheet \cite{mesa}, but how much larger? Equating the folding and kinetic energies requires $p_c$ of at least $2 \times 10^{-6} N/m$ i.e. $B>10^{-16}$ J ---a suspiciously larger value. \omitt{A higher bending stiffness also means a larger wave length since $\lambda=2\pi [B/(\rho g)]^{1/4}$. Unless motion is for some reason halted before self contact, a larger wave length leads to a larger displacement, much more than is observed. \no{Maybe delete the last sentence? We get $\lambda$ that is too large by using the value we're currently using as well}\tw{\tt ok.  the sentence seems to add un-needed qualifications}}

We do not see how to resolve the kinetic discrepancy discussed here without departing qualitatively from the wrinkle-to-fold model of Ref. \cite{LukaCerda}.  To resolve the discrepancy, either the kinetic energy must be smaller than our conservative bounds  or the driving force must be stronger than our bounds.  The former possibility seems unlikely.  Our estimates for the kinetic energy are based on direct observations of the motion, together with simple and unquestioned hydrodynamics. However, our account of the driving force depends explicitly on the mechanical properties of a folding monolayer leading to the pressure $p_c$ that we estimated.  Perhaps this mechanical picture is wrong.  What other forces are strong enough to trigger the observed motion?  One possible force is the force of adhesion between two folds that touch.  These forces are comparable to the surface tension of the fluid and are thus many times larger in magnitude than the  pressure $p_c$. If adhesion forces are responsible, then the observed jerks must take place only after the folds have touched.  This leaves unexplained how the folds came to touch and what causes the jerking to stop.  A second way to have a larger force is to abandon our picture of a simple molecular monolayer.  The observed jerks occur when many folds are already present.  The out-of-plane structure from these prior folds could well impart great rigidity to the surface layer and increase its buckling pressure by a large factor.  If such structures were important, it would qualitatively alter our picture of how the jerks occur.  There would no longer be a clear connection to the simple wrinkle-to-fold model that gave a plausible account of the jerk displacements.

The force responsible for jerks in the monolayers of Ref.~\cite{jpc06} may well be important in a broader context.  The wrinkle-fold transition has been implicated in a broader class of nanoscale systems: nanoparticle trilayers \cite{Binuhua} and single-component lipid monolayers \cite{KYLeeReview}.  The various energies in these systems are different, and the motion is also somewhat different from that of Ref. \cite{jpc06}.  Nevertheless, the discrepancy shown above may also apply to these other systems.  It is also possible that the force responsible for jerks affects structure as well as dynamics.  That is, the magnitude of a fold may be governed by other characteristic lengths than the wrinkle wavelength $\lambda$.  Still, the incipient instability might be due to the wrinkle-fold mechanism while the rapid subsequent motion is controlled by adhesion or some other force.

This work highlights the distinctive dynamics of thin solid sheets on fluids in general.  Equation (\ref{power32}) applies whenever such a sheet is accelerated by an unbalanced force, whether folding occurs or not. The equations imply a distinctive form of acceleration and a distinctive partitioning of the supplied work into kinetic energy and dissipation.

\section{Conclusion}
\label{conclusion}

The wrinkle-to-fold mechanism \cite{LukaCerda} for thin film buckling has allowed a new avenue for understanding buckling phenomena in a range of nanoscale systems.  Its successes in explaining structural aspects of folding have led us to apply it to the well-studied dynamics of monolayer jerks.  For these dynamic phenomena, our study indicates that other forces are at play.  Supporting this conclusion are experiments on lipid-coated micro-bubbles, where it is clear that gravity, an important ingredient of the folding mechanism, plays no role \cite{Longo, Kwan}. These other forces must be much stronger than those previously considered.  The necessity of such forces underscores the remarkable nature of monolayer jerks.  It also raises the importance of understanding these jerks.

\begin{acknowledgments}
We are grateful to Stuart Rice, Benny Davidovitch, Luka Pocivavsek, and Philippe Guyot-Sionnest for useful discussions.  This work was supported by the US--Israel
 Binational Science Foundation (Grant No. 2006076), and the University of Chicago MRSEC program of the NSF under Award Number DMR 0820054.
\end{acknowledgments}




\begin{thebibliography}{99}
\bibitem{bending_factor}
This $30$ factor is not to be confused with the energy difference mentioned in the abstract. In particular changing our choice of bending stiffness would not resolve the energy discrepancy noted in the abstract.  It would worsen this discrepancy.
\end{thebibliography}
\end{document}